\documentclass[english,aps,prd,showpacs,floatfix,nolongbibliography, nofootinbib, preprint,superscriptaddress]{revtex4-2}
\usepackage{parskip}
\usepackage{physics}
\usepackage{amsmath}
\usepackage{amssymb}
\usepackage[dvipsnames]{xcolor}
\usepackage{array}
\usepackage{longtable}
\usepackage{multirow}
\usepackage{makecell}
\usepackage{diagbox}
\usepackage{comment}
\usepackage{graphicx}
\usepackage{amsfonts}
\usepackage{bm}
\usepackage{slashed}
\usepackage{dsfont}
\usepackage{mathtools}
\usepackage[compat=1.1.0]{tikz-feynman}
\usepackage{mathrsfs}
\usepackage{xparse}
\usepackage{extarrows}
\usepackage{enumitem}
\usepackage{empheq}
\usepackage[T1]{fontenc}
\usepackage[scr=boondox]{mathalfa}

\usepackage[caption=false]{subfig}

\usepackage{simpler-wick}

\allowdisplaybreaks

\newcommand\MTkillspecial[1]{
	\bgroup
	\catcode`\&=9
	\let\\\relax%
	\scantokens{#1}%
	\egroup
	}
\DeclarePairedDelimiter\BraceM\{\}
\reDeclarePairedDelimiterInnerWrapper\BraceM{star}{
	\mathopen{#1\vphantom{\MTkillspecial{#2}}\kern-\nulldelimiterspace\right.}
	#2
	\mathclose{\left.\kern-\nulldelimiterspace\vphantom{\MTkillspecial{#2}}#3}
	}
\DeclarePairedDelimiter\bracketM{[}{]}
\reDeclarePairedDelimiterInnerWrapper\bracketM{star}{
	\mathopen{#1\vphantom{\MTkillspecial{#2}}\kern-\nulldelimiterspace\right.}
	#2
	\mathclose{\left.\kern-\nulldelimiterspace\vphantom{\MTkillspecial{#2}}#3}
	}

\usepackage{ifthen}

\newcommand{\mmd}[2][d]{\ifthenelse{\equal{#1}{1}}{\frac{\dd {#2}}{2\pi}}{\frac{\dd^{#1}{#2}}{(2\pi)^{#1}}}}

\makeatletter
\newcommand{\pushright}[1]{\ifmeasuring@#1\else\omit\hfill$\displaystyle#1$\fi\ignorespaces}
\newcommand{\pushleft}[1]{\ifmeasuring@#1\else\omit$\displaystyle#1$\hfill\fi\ignorespaces}
\makeatother

\NewDocumentCommand\NL{s}{%
  \IfBooleanTF#1%
    {\notag\\\times}
    {\notag\\}
}

\usepackage{shellesc}
\usetikzlibrary{external}

\tikzfeynmanset{
	Eikonal/.style={
		/tikz/draw=none,
		/tikz/decoration={name=none},
		/tikz/postaction={
			/tikz/draw,
			/tikz/double distance=2pt,
		},
	},
	half right/.append style={
		/tikz/looseness=2
	}, 
	half left/.append style={
		/tikz/looseness=2
	}, 
	/tikzfeynman/warn luatex=false
}

\usepackage{hyperref}

\graphicspath{{figs/}}

\begin{document}

\hypersetup{pageanchor=false}
\title{Impact of high invariant-mass Drell-Yan forward-backward asymmetry measurements on SMEFT fits}
\author{Radja Boughezal}
\email{rboughezal@anl.gov}
\affiliation{High Energy Physics Division, Argonne National Laboratory, Argonne, IL 60439, USA}
\author{Yingsheng Huang}
\email{yingsheng.huang@northwestern.edu}
\author{Frank Petriello}
\email{f-petriello@northwestern.edu}
\affiliation{High Energy Physics Division, Argonne National Laboratory, Argonne, IL 60439, USA}
\affiliation{Department of Physics \& Astronomy,
Northwestern University, Evanston, IL 60208, USA}

\begin{abstract}

	 We study the impact of LHC forward-backward asymmetry (AFB) measurements at high invariant-mass in the Drell-Yan process on probes of semileptonic four-fermion operators in the Standard Model effective field theory (SMEFT). In particular, we study whether AFB measurements can resolve degeneracies in the Wilson coefficient parameter space that appear when considering invariant-mass and rapidity measurements alone. We perform detailed fits of the available high-energy and high-luminosity ATLAS and CMS data for both invariant-mass distributions and AFB. While each type of measurement separately exhibits degeneracies, combining them removes these blind spots in some cases. In other situations it does not, highlighting the importance of incorporating future datasets from other experiments to fully explore this sector of the SMEFT. We investigate the impact of contributions quadratic in the Wilson coefficients on the description of Drell-Yan data and discuss when such terms are important in joint fits of the AFB and invariant-mass data. 
	 
\end{abstract}

\maketitle
\newpage

\hypersetup{pageanchor=true}

\section{Introduction}

The Standard Model (SM) successfully describes the fundamental interactions of the Universe. However, it cannot be the ultimate theory of nature as it fails to account for several observed phenomena. It does not include neutrino masses, nor does it contain a mechanism to generate the observed baryon-antibaryon asymmetry. Furthermore, it does not contain a description of dark matter nor dark energy. Experiments and observations across a broad spectrum of energies are searching for explanations of these outstanding issues. High-energy colliders such as the Large Hadron Collider (LHC) are actively testing the SM and searching for new physics at the TeV scale, but there are currently no conclusive discrepancies between SM predictions and experimental measurements. 

The current experimental landscape motivates indirect searches for heavy new physics beyond the direct reach of current collider energies. The SM effective field theory (SMEFT) is a framework for studying such heavy new physics effects. The SMEFT is an effective field theory extension of the SM containing only SM fields that preserves the gauge symmetries of the SM. New physics effects are encoded in a series of higher-dimensional operators, suppressed by a characteristic energy scale $\Lambda$, below which new physics degrees of freedom are integrated out. The complete operator bases of the SMEFT at dimension 6 and dimension 8 are known~\cite{Arzt:1994gp,Grzadkowski2010,Murphy2020,Li:2020gnx}. Odd-dimensional operators violate lepton {and baryon} number and are not considered in this work. Because of its model-independent nature, the SMEFT framework is particularly useful for investigating the effects of new physics on precision tests of the SM. There is an ongoing effort to analyze increasing amounts of available data within the SMEFT framework~\cite{Han:2004az,Pomarol:2013zra,daSilvaAlmeida:2018iqo,Aguilar-Saavedra:2018ksv,Falkowski:2019hvp,Ellis:2020unq,Ethier:2021bye,Crivellin:2021rbf}.

Our goal in this work is to study the impact of LHC Drell-Yan data on the allowed ranges of the Wilson coefficients in the SMEFT.  Drell-Yan is the natural process in which to probe the semileptonic four-fermion sector of the SMEFT. Although low-energy processes can play an important role in constraining these operators in some scenarios~\cite{Falkowski:2017pss,Boughezal:2021kla}, in general, these constraints are weaker than those provided by Drell-Yan at the LHC. Previous work has shown that the Drell-Yan process is potentially sensitive to operators at the dimension-8 level~\cite{Boughezal2021} and that measurements of the transverse momentum distribution can help distinguish between possible ultraviolet completions~\cite{Boughezal:2022nof}. The QCD and electroweak corrections to the SMEFT contributions are known up to next-to-leading order (NLO)~\cite{Dawson:2018dxp,Dawson:2021ofa}. Global fits of the available high invariant-mass distributions have been performed~\cite{Allwicher:2022gkm}. However, the Drell-Yan invariant-mass distributions can constrain only a limited number of combinations of four-fermion Wilson coefficients, and large swaths of the available parameter space cannot be probed~\cite{Alte:2018xgc,Boughezal:2020uwq}. Future colliders such as the electron-ion collider (EIC) can remove these degeneracies~\cite{Boughezal:2020uwq}.

We study here how measurements of the high invariant-mass forward-backward asymmetry (AFB) in Drell-Yan, in combination with invariant-mass distribution measurements, can probe these unconstrained four-fermion semileptonic directions in the SMEFT parameter space. We perform a global fit of available results from both ATLAS and CMS, and explain in detail our treatment of the available data. We find that the inclusion of AFB data can provide strong, multi-TeV constraints on semileptonic four-fermion parameters, and can remove some parameter degeneracies found in previous work. However, even the combination of invariant-mass and AFB data is blind to certain combinations of Wilson coefficients, highlighting the importance of future data from an EIC and elsewhere. We also show that including both AFB and invariant-mass data can help reduce the impact of quadratic dimension-6 terms in the SMEFT expansion. We discuss in what cases the quadratic terms can be neglected.

Our paper is organized as follows. In Sec.~\ref{sec:smeft}, we first review the four-fermion sector of SMEFT relevant to our analysis. In Sec.~\ref{sec:fit}, we describe our calculational framework and methodology for performing fits of the Drell-Yan data to the SMEFT. In particular, we demonstrate how to include existing AFB measurements in our fit. Since some of the available data are not fully unfolded this involves some assumptions, and we show that our analysis is relatively insensitive to this uncertainty. In Sec.~\ref{sec:results}, we present the results of the fits to the LHC data. Finally, we conclude in Sec.~\ref{sec:conclusion}.

\section{Review of the SMEFT\label{sec:smeft}}

In this section we review the aspects of the SMEFT relevant for our analysis. The SMEFT is an effective field theory extension of the SM in which new physics effects are encoded in a series of higher-dimensional operators. It features an expansion in energy over a dimensionful parameter $\Lambda$, the scale at which new particles are expected to appear. In the expansion in $1/\Lambda$ we keep operators only up to dimension-6 and neglect {both lepton and baryon} number-violating operators of odd dimension. Our SMEFT Lagrangian reads
\begin{equation}
	{\cal L} = {\cal L}_\textrm{SM}+ \frac{1}{\Lambda^2} \sum_i C^{(6)}_{i} {\cal
		O}^{(6)}_{i} 		
		+ \cdots,
\end{equation}
where the ellipsis denotes higher-dimensional operators, and the $C_i$ are dimensionless Wilson coefficients. In our numerical results later we set $\Lambda = 4$ TeV, significantly above the lower limits of the invariant-mass bins for all datasets in our study.

Operators relevant for the Drell-Yan process can be categorized as discussed in~\cite{Boughezal2021}. At dimension 6 at tree level, there are three types of operators: vertex corrections to the gauge boson couplings to fermions, dipole operators that couple fermions to gauge bosons, and semileptonic four-fermion operators. Contributions of the vertex corrections scale with energy as $\mathcal{O}(v^2/\Lambda^2)$, where $v$ is the Higgs vacuum expectation value. We assume minimal flavor violation (MFV) for the Wilson coefficients. As a result, {the scalar four-fermion and dipole operators for first and second generation fermions which are most relevant for our analysis }are proportional to the SM Yukawa couplings, which can be safely ignored. {The structure of the SMEFT operators in MFV is summarized in~\cite{Aoude2020}, where the Yukawa-coupling suppression is shown explicitly.} The vectorlike four-fermion operators scale as $\mathcal{O}(s/\Lambda^2)$. These four-fermion operators are the most relevant for our analysis at high energies, and we will focus on them in the following. We detail all the operators relevant for our analysis in Table~\ref{tab:ffops}. We note that corrections to the $ffZ$ vertex can also shift the Drell-Yan cross section. These are better probed through AFB measurements in on-shell $Z$-boson production~\cite{Breso-Pla:2021qoe}, and their effects tend to be smaller than those from four-fermion operators in high-mass Drell-Yan production~\cite{Boughezal2021}. We note that one-loop corrections to the Drell-Yan process in the SMEFT have been considered~\cite{Dawson:2018dxp}.

Previous fits to the Drell-Yan invariant-mass data have shown that both quadratic dimension-6 effects and certain classes of dimension-8 operators can significantly shift the constraints obtained on the dimension-6 Wilson coefficients~\cite{Boughezal2021}. We study later the inclusion of quadratic dimension-6 terms in our results. We find that simultaneously fitting both invariant-mass and forward-backward asymmetry data can significantly reduce the impact of quadratic terms, improving the convergence of the SMEFT expansion when applied to these data. In some cases the joint fit with quadratic effects is numerically very similar to the linear fit, further showing the importance of including both datasets in the analyses of Drell-Yan data. This is shown explicitly in our numerical results section. We discuss later the conditions required for the quadratic and linear fits to coincide.

\setlength{\tabcolsep}{15pt}
\begin{table}[htbp]
	\centering
	\begin{tabular}{lclc}
	\hline
	\hline
	${\cal O}_{lq}^{(1)}$ & $(\bar{l}\gamma^{\mu} l) (\bar{q}\gamma_{\mu} q)$ &  ${\cal O}_{lu}$ & $(\bar{l}\gamma^{\mu} l) (\bar{u}\gamma_{\mu}  u)$ 
	\\
	${\cal O}_{lq}^{(3)}$ & $(\bar{l}\gamma^{\mu} \tau^I l)
											(\bar{q}\gamma_{\mu} \tau^I l q)$ & ${\cal O}_{ld}$ & $(\bar{l}\gamma^{\mu} l) (\bar{d}\gamma_{\mu}  d)$   
	\\
	${\cal O}_{eu}$ & $(\bar{e}\gamma^{\mu} e) (\bar{u}\gamma_{\mu}  u)$
																																			& ${\cal O}_{qe}$ & $(\bar{q}\gamma^{\mu} q) (\bar{e}\gamma_{\mu}  e)$             
	\\
	${\cal O}_{ed}$ & $(\bar{e}\gamma^{\mu} e) (\bar{d}\gamma_{\mu}  d)$
																																			& &
	\\
	\hline
	\hline
	\end{tabular}
	\caption{Dimension-6 four-fermion operators contributing to Drell-Yan at leading order in the coupling constants. $q$ and $l$ denote quark and lepton doublets, respectively, while $u$, $d$ and $e$ denote right-handed singlets for up-type quarks, down-type quarks and leptons, respectively. $\tau$ is the SU(2) Pauli matrix.
	\label{tab:ffops}}
\end{table}

\section{Description of the analysis framework \label{sec:fit}}

In this section we describe our procedure for performing fits of $\dv*{\sigma}{m}$ and AFB measurements of the neutral current Drell-Yan process at the LHC within the SMEFT framework. We describe the details of our theoretical calculation. We discuss our treatment of the experimental data and the assumptions we make in dealing with some datasets for which the full experimental error correlations are unavailable.

\subsection{Experimental data}

For our analysis, we demand that the LHC datasets used feature high dilepton invariant-mass and high luminosity. The former assumption enhances the impact of SMEFT four-fermion operators, while the latter ensures enough statistics for the SMEFT Wilson coefficients to be well probed. These requirements lead us to the following four datasets at the LHC: two single-differential cross section ($\dv*{\sigma}{m}$) measurements at $8$~\cite{ATLAS:2016gic} and $13\ \mathrm{TeV}$~\cite{CMS:2021ctt}, and two AFB measurements at $8$~\cite{CMS:2016bil} and $13\ \mathrm{TeV}$~\cite{CMS:2022uul}, respectively. These datasets are chosen specifically for their high luminosities and high dilepton invariant-mass (up to $7\ \mathrm{TeV}$). In our study, we use only high-$m_{ll}$ bins with $\hat{s}\gg M_Z^2$. The exact binnings for each dataset are listed in the Appendix. The relevant information for each dataset is given in Table~\ref{tab:datasets}, where we also label them for future reference. All measurements are unfolded, with the electron and muon channels combined, with the exception of the $13 \ \mathrm{TeV}$ CMS measurement of $\dv*{\sigma}{m}$. We separately treat the electron and muon channels in this measurement, as they have different integrated luminosities and $m_{ll}$ binnings. We also directly use event yields in this measurement, as cross sections are not publicly available. 

{\setlength{\tabcolsep}{6pt}
\begin{table}[htbp]
	\centering
	\begin{tabular}{c|lccccc}
		\hline
		\hline
		No. & Experiment & $\sqrt{s}$          & Measurement       & Luminosity               & $m_{ll}^\textrm{low}$       & Ref.                 \\ \hline
		I   & ATLAS      & $8 \ \mathrm{TeV}$  & $\dv*{\sigma}{m}$ & $20.3\ \mathrm{fb}^{-1}$ & $116$-$1000 \ \mathrm{GeV}$ & \cite{ATLAS:2016gic}     \\
		II   & CMS        & $13 \ \mathrm{TeV}$ & $\dv*{\sigma}{m}$ & \makecell{$137 \ \mathrm{fb}^{-1}\ (ee)$ \\$140 \ \mathrm{fb}^{-1}\ (\mu\mu)$} &  \makecell{$200$-$2210 \ \mathrm{GeV}\ (ee)$\\$210$-$2290 \ \mathrm{GeV}\ (\mu\mu)$}     & \cite{CMS:2021ctt}     \\
		III   & CMS        & $8 \ \mathrm{TeV}$  & $A_\mathrm{FB}^*$   & $19.7 \ \mathrm{fb}^{-1}$ &  $120$-$500 \ \mathrm{GeV}$             & \cite{CMS:2016bil}      \\
		IV   & CMS        & $13 \ \mathrm{TeV}$ & $A_\mathrm{FB}$   & $138 \ \mathrm{fb}^{-1}$ &  $170$-$1000 \ \mathrm{GeV}$             & \cite{CMS:2022uul} \\ 
		\hline
		\hline
	\end{tabular}
	\caption{Summary of the Drell-Yan datasets used in this analysis. The first column indexes all datasets numerically. $m_{ll}^\textrm{low}$ denotes the range of the lower edges of the dilepton invariant-mass used in this work. More details regarding the binning is given in the Appendix. }
	\label{tab:datasets}
\end{table}}
%

\subsection{Theoretical calculations}

We compute the SM cross sections using the {MCFM} program~\cite{Campbell:2019dru}. For each dataset, we compute the SM cross section to NLO in QCD with next-to-next-to-leading order NNPDF 3.1 parton distribution functions (PDFs)~\cite{NNPDF:2017mvq}. We have checked that the NNLO QCD corrections to both the invariant-mass and AFB distributions are small, leading to a correction that is less than $2\%$ for most energies. To avoid unnecessary computational expense we neglect them in our fit. Since we are probing the high-energy region of the Drell-Yan data, electroweak Sudakov logarithms become important and can lead to shape differences that mimic SMEFT effects. We include these corrections and incorporate them by multiplying each bin of the QCD-corrected cross section with the ratio of the NLO electroweak result divided by the LO cross section.

\begin{figure}[htbp]
	\centering
	\includegraphics[width=0.49\textwidth]{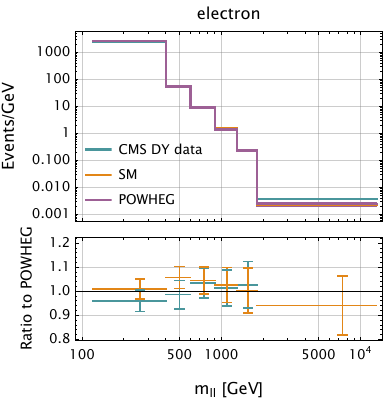}
	\includegraphics[width=0.49\textwidth]{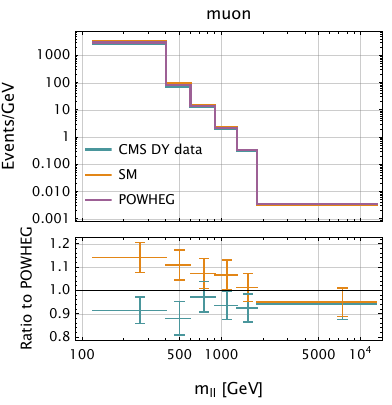}
	\caption{Event yields in the electron (left) and muon (right) channels for the $13\ \mathrm{TeV}$ dataset II, as detailed in Table\ref{tab:datasets}. The green lines show the observed total event yields minus all non-Drell-Yan backgrounds. The orange lines show our SM predictions with electroweak Sudakov corrections. The purple line shows the {POWHEG} estimate for the Drell-Yan background.
	The lower inset shows the ratio to the Drell-Yan background estimations in Ref.~\cite{CMS:2021ctt}. The error bars represent uncertainties from the {POWHEG} estimates. }
	\label{fig:eventyields}
\end{figure}

For dataset II, we need the predicted event yields to compare with experimental data. By interpolating  the "Acceptance$\times$Efficiency" data in {HEPData}~\cite{https://doi.org/10.17182/hepdata.101186.v2}, along with the luminosities provided in Table~\ref{tab:datasets}, we obtain the SM event yields. We compare our results with the {POWHEG}-produced Drell-Yan background estimations provided in~\cite{CMS:2021ctt}. In Fig.~\ref{fig:eventyields}, we show our predicted event yields for the electron and muon channels. For the experimental values, we take the observed total event yields and subtract the non-Drell-Yan backgrounds. The predicted event yields are, in general, in good agreement with the {POWHEG} background estimations, except for the muon channel in the low-$m_{ll}$ region. We note that the electroweak Sudakov logarithms have a significant effect ${\cal O}(10\%)$ in the higher invariant-mass bins. In order to take advantage of the high-$m_{ll}$ region probed by dataset II, we choose the finer binnings as shown in Table~\ref{tab:binning} instead of the binnings in Fig.~\ref{fig:eventyields}. One drawback of this choice is that the dataset lacks non-Drell-Yan backgrounds for the finer binning. Non-Drell-Yan backgrounds are not negligible in size, and in order to quantitatively measure the deviation of SM from the observed total event yields, it is necessary to have both the Drell-Yan and non-Drell-Yan contributions. However, our theoretical calculation covers only the Drell-Yan contributions. Therefore, without the non-Drell-Yan backgrounds from {CMS}, we cannot make direct comparisons between our theory predictions and the observed data. To work around this issue, we replace our own predictions of the SM with the {POWHEG} background estimates as our central values. As shown in Fig.~\ref{fig:eventyields}, since our predictions closely align with the {POWHEG} estimates, this choice has a negligible impact. The theoretical uncertainties for dataset II are still generated from our own predictions.

For datasets III and IV, a further step is needed to obtain the forward-backward asymmetry. The definition of AFB is given by
\begin{align}
	A_\mathrm{FB} = \frac{\sigma_{\mathrm{F}}-\sigma_{\mathrm{B}}}{\sigma_{\mathrm{F}}+\sigma_{\mathrm{B}}}.
\end{align}
$\sigma_{\mathrm{F}}$ ($\sigma_{\mathrm{B}}$) is the forward (backward) cross section, defined by $\cos\theta^*>0$ ($\cos\theta^*<0$), where $\theta^*$ is the angle between the negatively charged lepton and the initial state quark in the dilepton center-of-mass (c.m.) frame~\cite{Collins:1977iv}. In general, $\theta^*$ is defined as
\begin{align}
	\cos \theta^*=\frac{2\left(P_1^{+} P_2^{-}-P_1^{-} P_2^{+}\right)}{\sqrt{Q^2\left(Q^2+Q_{\mathrm{T}}^2\right)}}
\end{align}
where $Q$ and $Q_T$ represent the four-momentum and the transverse momentum of the dilepton system, respectively, $P_1$ ($P_2$) is the four-momentum of the negatively (positively) charged lepton with $P_i^{\pm}=(E_i\pm P_i^z)/\sqrt{2}$, and $E_i$ is the energy of lepton $i$. This definition, however, depends on the identification of the quark direction, which is not experimentally observable. In dataset III, the quark direction is approximately identified using the sign of the $z$-component of dilepton momentum:
\begin{align}
	\cos \theta_R= \frac{\abs{Q^z}}{Q^z}\cos \theta^*. 
\end{align}
The accuracy of this approximation relies upon the stiffness of the valence-quark distributions compared to the antiquark distributions. CMS takes a different approach in dataset IV. The quark direction is captured in Monte Carlo simulations, where one has access to the partonic c.m. frame. To mimic this, we manually pick out the quark and antiquark directions in our prediction.\footnote{In the case of gluon-initiated processes, we assign the ``true'' quark direction in a manner consistent with the cancellation of collinear singularities in the perturbative calculation.} In this work, we denote the former definition of AFB as $A_\mathrm{FB}^*$, while the latter, often referred to as the ``true'' AFB in the literature~\cite{CMS:2022uul,Accomando:2015cfa}, is denoted as $A_\mathrm{FB}$. We comment later on several aspects of $A_\mathrm{FB}$.

\begin{figure}[!htbp]
	\centering
	\includegraphics[width=0.49\linewidth]{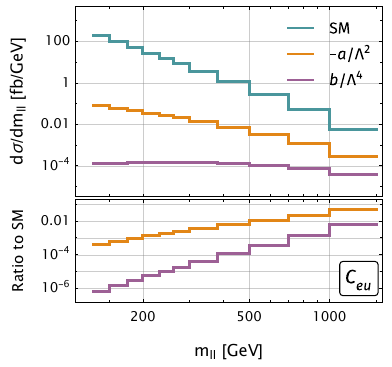}
	\includegraphics[width=0.49\linewidth]{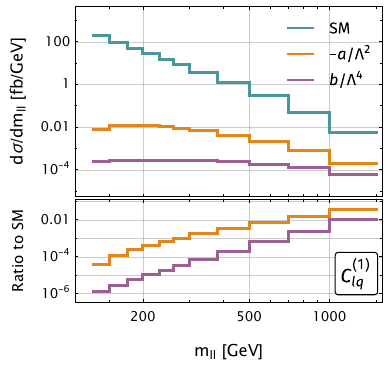}
	\caption{The linear and quadratic SMEFT contributions to the cross section for the Wilson coefficients $C_{eu}$ (left) and $C_{lq}^{(1)}$ (right), respectively. The orange and purple lines represent the linear ($a/\Lambda^2$) and quadratic ($b/\Lambda^4$) SMEFT contributions with $C=1$ and $\Lambda=4\ \mathrm{TeV}$, while the green line shows the SM contribution. The binning and fiducial cuts are those of dataset I in Table~\ref{tab:datasets}. \label{fig:dsigmaSMEFT}}
\end{figure}
\begin{figure}[!htbp]
	\centering
	\includegraphics[width=0.49\linewidth]{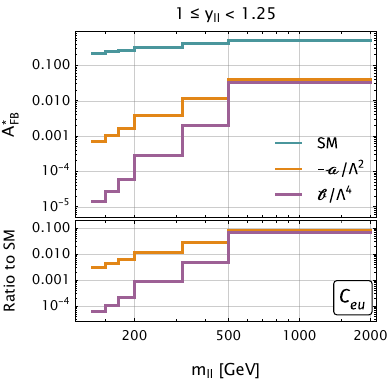}
	\includegraphics[width=0.49\linewidth]{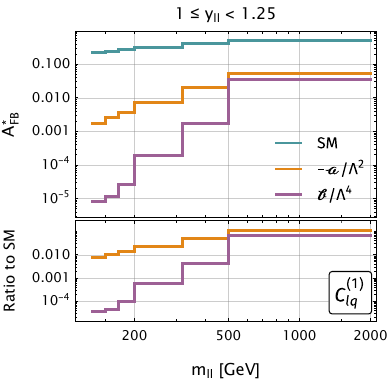}
	\caption{The linear and quadratic SMEFT contributions to $A_{\mathrm{FB}}^*$ for the Wilson coefficients $C_{eu}$ (left) and $C_{lq}^{(1)}$ (right), respectively. The orange and purple lines represent the linear ($\mathscr{a}/\Lambda^2$) and quadratic ($\mathscr{b}/\Lambda^4$) SMEFT contributions with $C=1$ and $\Lambda=4\ \mathrm{TeV}$, while the green line shows the SM contribution. The binning, fiducial cuts and the definition of AFB are those of dataset III in Table~\ref{tab:datasets}. \label{fig:AFBSMEFT}}
\end{figure}

For the SMEFT contributions to the cross sections, we work at LO in QCD. The total cross section contains both the SM contribution and SMEFT contributions that are linear and quadratic in dimension-6 Wilson coefficients $C_i^{(6)}$:
\begin{align}
	\dd\sigma=\dd\sigma_{\mathrm{S M}}+\sum_i\frac{C_i^{(6)}}{\Lambda^2}a_i^{(6)}+\sum_{ij}\frac{C_i^{(6)}C_j^{(6)}}{\Lambda^4}b_{ij}^{(6)}.
	\label{eq:xsection}
\end{align}
The $a$ and $b$ terms signify the linear and quadratic SMEFT contributions from dimension-6 operators, respectively.  
Additionally, we expand the full AFB up to $\mathcal{O}(1/\Lambda^{4})$: 
{\begin{align}
	A_\mathrm{FB}&=A_\mathrm{FB}^{\mathrm{SM}}+\sum_i\frac{C_i^{(6)}}{\Lambda^2}\;\mathscr{a}_i^{(6)}+\sum_{ij}\frac{C_i^{(6)}C_j^{(6)}}{\Lambda^4}\;\mathscr{b}_{ij}^{(6)}
	\notag\\&
	=A_\mathrm{FB}^{\mathrm{SM}}+\sum_i\frac{C_i^{(6)}}{\Lambda^2}\frac{\sigma_{\mathrm{S M}} \Delta a_i^{(6)}-a_i^{(6)} \Delta \sigma_{\mathrm{S M}}}{\sigma_{\mathrm{S M}}^2}
	\notag\\&
	+\sum_{ij}\frac{C_i^{(6)}C_j^{(6)}}{\Lambda^4}\frac{\pqty{a_i^{(6)}}^2\Delta\sigma_{\mathrm{S M}}-a_i^{(6)}\Delta a_i^{(6)}\Delta\sigma_{\mathrm{S M}} -b_{ij}^{(6)}\sigma_{\mathrm{S M}}\Delta\sigma_{\mathrm{S M}}+\Delta b_{ij}^{(6)} \sigma_{\mathrm{S M}}^2}{\sigma_{\mathrm{S M}}^3}
	 +\mathcal{O}(1/\Lambda^{6}) , 
	\label{eq:afb}
\end{align}}
where $\sigma\equiv\sigma_F+\sigma_B$ and $\Delta \sigma\equiv\sigma_F-\sigma_B$, {while $\mathscr{a}$ and $\mathscr{b}$ terms correspond to the linear and quadratic SMEFT contributions to AFB, respectively. }{In Fig.~\ref{fig:dsigmaSMEFT} and \ref{fig:AFBSMEFT}, we show the linear and quadratic SMEFT contributions to the cross section and AFB for both $C_{eu}$ and $C_{lq}^{(1)}$. }In the following section, we keep terms up to $\mathcal{O}(1/\Lambda^2)$ in the SMEFT contributions. We study the impact of quadratic terms in the last section.

\subsection{Uncertainties and $\chi^2$ tests}

\begin{figure}[htbp]
	\centering
	\includegraphics[width=0.49\textwidth]{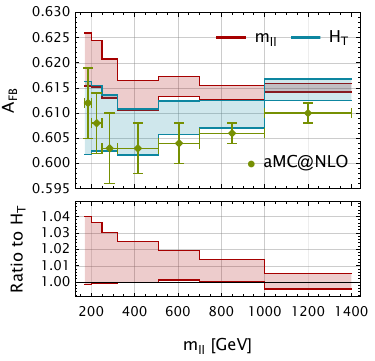}
	\includegraphics[width=0.49\textwidth]{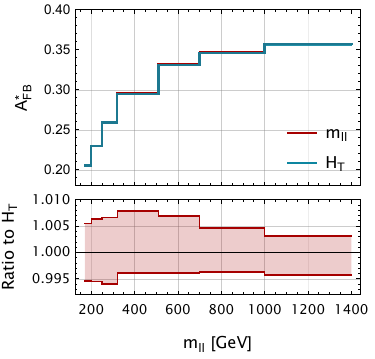}
	\caption{Left panel: the ``true'' forward-backward asymmetry $A_{\mathrm{FB}}$ for dataset IV (labeled in Table~\ref{tab:datasets}) with dynamic scale $\mu_0=m_{ll}$ (red) and $\mu_0=H_T$ (blue). The bands represent the range of scale variation ($1/2 \leq \mu_{R,F}/\mu_0 \leq 2, \;\; 1/2 \leq \mu_{R}/\mu_F \leq 2$) for both scale choices. The {aMC@NLO} simulation in Ref.~\cite{CMS:2022uul} is shown by the green points. Right panel: the same comparison for $A^{*}_{\mathrm{FB}}$. }
	\label{fig:dataset3}
\end{figure}

To calculate the PDF uncertainties, we follow the standard procedure for Monte Carlo replica sets~\cite{NNPDF:2017mvq}. We correlate these errors across all four datasets. To estimate the uncertainty arising from higher-order QCD corrections, we set the renormalization and factorization scales to a central value $\mu_0$ and vary them around this value in an uncorrelated way according to 
\begin{align}
	\frac{1}{2} \leq \mu_{R,F}/\mu_0 \leq 2, \;\; \frac{1}{2} \leq \mu_{R}/\mu_F \leq 2.
\end{align}
We find the largest variation in either direction within this range and form a symmetric scale uncertainty from it. We choose the dynamic scale for all datasets to be $\mu_0=m_{ll}$, except for dataset IV. We note that in the special case of dataset IV, the ``true'' AFB with dynamic scale $\mu_0=m_{ll}$ significantly deviates from the {aMC@NLO} simulation in Ref.~\cite{CMS:2022uul} as shown in Fig.~\ref{fig:dataset3}. Therefore, we also consider the case of $\mu_0=H_T$, where $H_T$ is the sum of transverse masses of all final state particles. We use $\mu_0=H_T$ as the central value, and vary $\mu_{R,F}$ around it in the same way as above. We combine the scale variations of both $\mu_0=m_{ll}$ and $\mu_0=H_T$. The largest variation from the central value $\mu_0=H_T$ is then used to form a symmetric scale uncertainty. We note that the poor perturbative behavior we find suggests that this definition of angle for AFB is not preferred, at least as far as the theory calculation is concerned. The same scale choice comparison is also shown for  $A^{*}_{\mathrm{FB}}$ in Fig.~\ref{fig:dataset3}. This definition of AFB leads to a greatly improved agreement between the two scale choices. While our implementation of $A_{\mathrm{FB}}$ leads to results in acceptable agreement with the {aMC@NLO} simulation in Ref.~\cite{CMS:2022uul}, it is unclear to us how to extend it to higher orders in the QCD perturbative expansion. Certainly if an improved theoretical description is a priority then the $A_\mathrm{FB}^*$ definition should be adopted. {The relative scale and PDF uncertainties for most bins are below $3\%$. The relative scale uncertainties of the $\dv*{\sigma}{m}$ datasets increase with $m_{ll}$, while the uncertainties for $A_{\mathrm{FB}}$ in dataset IV decrease with $m_{ll}$. The uncertainties for $A_{\mathrm{FB}}^*$ in dataset III are tiny and show no significant dependence on $m_{ll}$. The relative PDF uncertainties of the $\dv*{\sigma}{m}$ datasets also increase with $m_{ll}$, but the uncertainties for $A_{\mathrm{FB}}$ and $A_{\mathrm{FB}}^*$ in datasets III and IV show no obvious dependence on $m_{ll}$. The experimental systematic uncertainties of all datasets increase with $m_{ll}$. }

We construct the experimental error matrices based on the published information from the CMS and ATLAS Collaborations. We assume there is no correlation between the datasets. For dataset I, we use the experimental errors provided by ATLAS. For the other datasets, since the information about correlated errors is not provided in the experimental papers, assumptions have to be made. 
For dataset II, we assume there are no correlations between the uncertainties of the different bins, and also between the different channels~\footnote{This treatment is consistent with Allwicher et al.~\cite{Allwicher:2022gkm,Allwicher:2022mcg}. }. 
For dataset III, there are two approaches we take: assuming no correlation between bins, and assuming the nearest off-diagonal elements in the systematic error matrix to be $15\%$ of the diagonal elements. We find little difference between these two assumptions in our fits.
For dataset IV, we also assume the uncertainties are uncorrelated between bins, motivated by our findings when trying different correlation assumptions for dataset III. 

Finally, we assemble the experimental and theoretical results to perform $\chi^2$ fits for both single datasets and globally. We use the following $\chi^2$ statistic to quantify the deviation of the SMEFT {prediction} from the {experimental data}: 
\begin{align}
	{\chi^2=\sum_{i,j}^{\textrm{\#of bins}}\frac{\left(\sigma_{i}^{\mathrm{exp}}-\sigma_{i}^{\mathrm{SMEFT}}\right)\left(\sigma_{j}^{\mathrm{exp}}-\sigma_{j}^{\mathrm{SMEFT}}\right)}{\Delta \sigma_{i j}^{2}},}
\end{align}
where $\Delta\sigma^2_{ij}$ signifies the error matrix composed of both theoretical and experimental uncertainties. We then extract the 68\% confidence level (C.L.) bounds of the Wilson coefficients based on $\chi^2$ fits. The values of $\chi^2/$d.o.f. with only SM contributions are listed in Table~\ref{tab:chi2}. We also compare the results with and without electroweak Sudakov logarithms. We find that the AFB measurements are minimally impacted by the electroweak Sudakov logarithms, while the fit to the invariant-mass distribution of dataset I improves with the inclusion of the electroweak Sudakov logarithms. In all cases we find good agreement between the SM and experiment.

\begin{table}[htbp]
	\centering
	\begin{tabular}{c|cccc}
		\hline
		\hline
		\backslashbox{$\chi^2$}{dataset} & I & II & III & IV \\
		\hline
		w/o Sudakov & 6.5/10 & - & 15/28 & 5.4/7 \\
		w/ Sudakov & 5.7/10 & 18.9/77 & 15/28 & 5.2/7 \\
		\hline
		\hline
	\end{tabular}
	\caption{The values of $\chi^2/$d.o.f. with only SM contributions for all four datasets, as labeled in Table~\ref{tab:datasets}. The first row is without electroweak Sudakov logarithms, and the second row is with electroweak Sudakov logarithms. The $\chi^2$ of dataset II is obtained from the {POWHEG} simulations in Ref.~\cite{CMS:2021ctt}. }
	\label{tab:chi2}
\end{table}

\section{Results\label{sec:results}}

In this section, we present the results of the $\chi^2$ fits. Since a major goal of our investigations is to study how combining invariant-mass and AFB data removes degeneracies in the space of Wilson coefficients, we choose four example combinations of nonzero Wilson coefficients to illustrate this point. These examples are adapted from~\cite{Boughezal:2020uwq}. To motivate our study below we recall that in the high-energy limit, the LO SMEFT corrections to the Drell-Yan partonic cross section can be written in the schematic form
\begin{equation}
\frac{\text{d}\sigma^x}{dm_{ll}^2 dY dc^{*}_{\theta}} \sim  \frac{1}{\Lambda^2}\frac{A_1^x \hat{u}^2+A_2^x\hat{t}^2}{\hat{s}^2}
	+\frac{1}{\Lambda^4} \left(B_1^x \hat{u}^2+B_2^x\hat{t}^2 \right).
\label{eq:DYcr}
\end{equation}
Here, $Y$ is the dilepton rapidity, $c^{*}_{\theta}$ is an abbreviation for the cosine of the lepton direction defined previously, and $\hat{s}$, $\hat{u}$ and $\hat{t}$ are the standard partonic Mandelstam invariants that can be written in the form
\begin{eqnarray}
\hat{t} &=& -\frac{\hat{s}}{2} (1-c^{*}_{\theta}), \nonumber \\
\hat{u} &=& -\frac{\hat{s}}{2} (1+c^{*}_{\theta}).
\end{eqnarray}
The coefficients $A_1^x$ and $A_2^x$ contain SM couplings and Wilson coefficients, and are linear in the Wilson coefficients. The coefficients $B_1^x$ and $B_2^x$ are quadratic in the dimension-6 Wilson coefficients. $x$ denotes the partonic channel, either up quark or down quark. For future reference, we note the functional dependences of these coefficients on the various Wilson coefficients, which we take from~\cite{Boughezal:2020uwq}:
\begin{eqnarray}
A_1^u &=& A_1^u(C_{eu},C_{lq}^{(1)},C_{lq}^{(3)}), \nonumber \\
A_1^d &=& A_1^d(C_{ed},C_{lq}^{(1)},C_{lq}^{(3)}), \nonumber \\
A_2^u &=& A_2^u(C_{lu},C_{qe}), \nonumber \\
A_2^d &=& A_2^d(C_{ld},C_{qe}).
\end{eqnarray} 
The $B_i^x$ depend upon the same Wilson coefficients as the corresponding $A_i^x$. In the high-energy limit  $\hat{s} \gg M_Z^2$ neither the $A$ nor the $B$ depend on any kinematic quantity. As shown in~\cite{Boughezal:2020uwq}, when expanded to linear order in the Wilson coefficients the invariant-mass distribution is only sensitive to the combination $A_1+A_2$, leading to blind spots in the parameter space. AFB is sensitive to $A_1-A_2$, making it possible to remove some of these degeneracies. We pick different nonzero Wilson coefficients that activate different combinations of $A_1$ and $A_2$ to see what can be learned from the AFB data. We note that it is, in general, not possible to simultaneously set the $A$ and $B$ structures to zero, and, therefore, the SMEFT contributions do not vanish if terms quadratic in the Wilson coefficients are retained. Their impact on the fits is discussed in the next section. {Other scenarios for the Wilson coefficients have been studied in the literature, such as the universal limit in which new physics can be parameterized completely in terms modifications of electroweak gauge boson propagators~\cite{Farina:2016rws}.} 

\subsection{Case I}

We begin with the case when $C_{eu}$, $C_{ed}$, and $C_{qe}$ are nonzero. This choice populates both $A_1$ and $A_2$ in Eq.~(\ref{eq:DYcr}): $C_{eu}$ and $C_{ed}$ contribute to $A_1^u$ and $A_1^d$, respectively, while $C_{qe}$ contributes to both $A_2^{u,d}$. In the high invariant-mass limit, after integrating over the angular variables, the SMEFT contributions to the invariant-mass distribution vanish under the following conditions~\cite{Boughezal:2020uwq}:
\begin{subequations}
	\begin{align}
		C_{q e} & =-C_{e u} \frac{Q_u e^2-g_Z^2 g_L^u g_R^e}{Q_u e^2-g_Z^2 g_R^e g_R^u} , \\
		C_{q e} & =-C_{e d} \frac{Q_d e^2-g_Z^2 g_L^d g_R^e}{Q_d e^2-g_Z^2 g_R^e g_R^d} ,
	\end{align}
\end{subequations}
where the SM left-handed and right-handed fermion couplings follow the convention of Ref.~\cite{Denner:1991kt}:
\begin{align}
	g_L^f=I_3^f-Q_f s_W^2, \quad g_R^f=-Q_f s_W^2 .
\end{align}
These conditions are simultaneously satisfied when 
\begin{align}
	C_{e d}^{(1)}\equiv C_{e d}=C_{e u} \frac{Q_u e^2-g_Z^2 g_L^u g_R^e}{Q_u e^2-g_Z^2 g_R^e g_R^u} \frac{Q_d e^2-g_Z^2 g_R^e g_R^d}{Q_d e^2-g_Z^2 g_L^d g_R^e}. 
	\label{eq:condition1}
\end{align}
This comes from setting $A_1^x+A_2^x$ to zero for both the up-quark and down-quark partonic channels. However, AFB does not vanish, suggesting that inclusion of these data can remove this degeneracy. We adopt a bottom-up perspective in this work and make no attempt to connect these parameter relations to an underlying ultraviolet model. We note that the vanishing of the SMEFT corrections occurs in the limit $\hat{s} \gg M_Z^2$, and is, therefore, approximate only. We allow $C_{qe}$ and $C_{eu}$ to vary subject to the constraint above. Fixing $C_{ed}$ allows us to visualize the bounds on a 2D plane and demonstrate the strong correlation between $C_{qe}$, $C_{eu}$ and $C_{ed}$. 

\begin{figure}[htbp]
	\centering
	\includegraphics[width=0.55\linewidth]{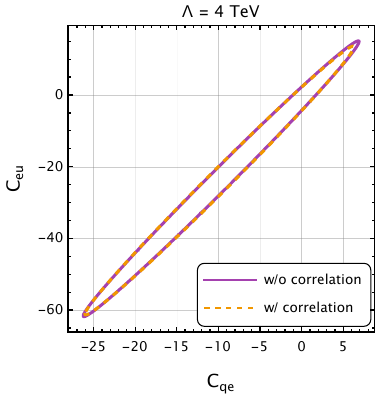}
	\caption{The comparison between assumptions for the experimental error matrices of dataset {III} (as labeled in Table~\ref{tab:datasets}). The ellipses represent the 68\% C.L. bounds of the Wilson coefficients $C_{e u}$ and $C_{q e}$ from the $\chi^2$ fits. The solid and dashed ellipses correspond to a diagonal experimental error matrix and an experimental error matrix with 15\% correlated errors between the nearest bins, respectively.}
	\label{fig:AFB8}
\end{figure}

We perform $\chi^2$ fits for each dataset separately and also for the combination of datasets, as described in Sec.~\ref{sec:fit}. For dataset III, we perform two fits: one with the assumption that the uncertainties are uncorrelated between bins, and the other with the assumption that the next-to-diagonal elements in the systematic error matrix are $15\%$ of the diagonal elements. We show in Fig.~\ref{fig:AFB8} the comparison between the two assumptions. The differences between them are small. We, therefore, choose the diagonal assumption for the rest of the analysis. 

\begin{figure}[htbp]
	\centering
	\includegraphics[width=0.55\linewidth]{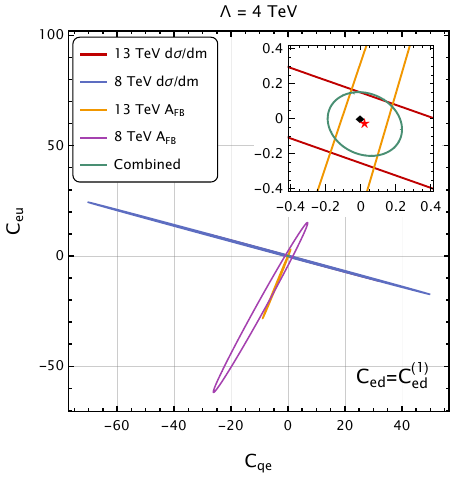}
	\caption{The 68\% C.L. bounds on the Wilson coefficients $C_{e u}$ and $C_{q e}$ from the $\chi^2$ fits (case I). The {red, purple} and orange ellipses correspond to the fits with dataset {II, III} and IV (as labeled in Table~\ref{tab:datasets}), respectively. The green ellipse corresponds to the combined fit with all four datasets. 
	The best-fit result is denoted with a red star, while the SM is denoted with a black diamond.}
	\label{fig:results:Cqe+Ceu}
\end{figure}

The 68\% C.L. ellipses of $C_{qe}$ and $C_{eu}$ are shown in Fig.~\ref{fig:results:Cqe+Ceu}. 
The result is exactly as expected: While the separate constraints from $\dv*{\sigma}{m}$ and AFB exhibit elongated ellipses indicating approximate degeneracies in the parameter space, these flat directions are almost orthogonal to each other. Consequently, the combined fit is able to break the degeneracies and constrain the individual Wilson coefficients. We note that the higher-luminosity 13 TeV datasets are significantly more constraining than the lower-luminosity 8 TeV datasets, as expected.


\subsection{Case II}

We next consider the case when $C_{eu}$, $C_{ed}$ and $C_{lq}^{(1)}$ are nonzero. This choice populates only the $A_1$ term in Eq.~(\ref{eq:DYcr}). In the high-invariant-mass limit, the SMEFT contributions vanish under the following conditions~\cite{Boughezal:2020uwq}:
\begin{subequations}
	\begin{align}
		& C_{l q}^{(1)}=-C_{e u} \frac{Q_u e^2-g_Z^2 g_R^u g_R^e}{Q_u e^2-g_Z^2 g_L^e g_L^u}  \\
		& C_{l q}^{(1)}=-C_{e d} \frac{Q_d e^2-g_Z^2 g_R^d g_R^e}{Q_d e^2-g_Z^2 g_L^e g_L^d} 
		\end{align}
\end{subequations}
These conditions are simultaneously satisfied when 
\begin{align}
	C_{e d}^{(2)}\equiv C_{e d}=C_{e u} \frac{Q_u e^2-g_Z^2 g_R^u g_R^e}{Q_u e^2-g_Z^2 g_L^e g_L^u} \frac{Q_d e^2-g_Z^2 g_L^e g_L^d}{Q_d e^2-g_Z^2 g_R^d g_R^e}. 
	\label{eq:condition2}
\end{align}
In this case we do not expect AFB to remove the degeneracy, since both AFB and the invariant-mass distribution probe the same combination of Wilson coefficients encoded in $A_1$.

\begin{figure}[htbp]
	\centering
	\includegraphics[width=0.55\linewidth]{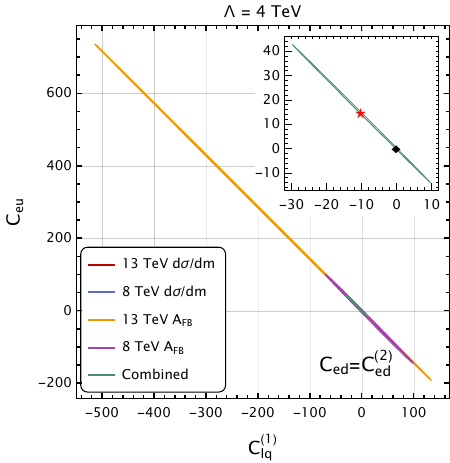}
	\caption{The 68\% C.L. bounds of the Wilson coefficients $C_{e u}$ and $C^{(1)}_{lq}$ from the $\chi^2$ fits (case II). The {blue, red, purple and orange} ellipses correspond to the fits with dataset {I, II, III} and IV (as labeled in Table~\ref{tab:datasets}), respectively. The green ellipse corresponds to the combined fit with all four datasets. The best-fit result is denoted with a red star, while the SM is denoted with a black diamond. }
	\label{fig:results:Clq1+Ceu}
\end{figure}
The 68\% C.L. ellipses of $C_{lq}^{(1)}$ and $C_{eu}$ are shown in Fig.~\ref{fig:results:Clq1+Ceu} after imposing Eq.~(\ref{eq:condition2}). As expected, the constraint ellipses from $\dv*{\sigma}{m}$ and AFB are now almost parallel to each other. Consequently, the combined fit is not able to resolve the flat directions. In this case, the AFB measurements do not improve the constraints on $C_{eu}$ and $C_{lq}^{(1)}$, and the combined fit closely resembles the $13\ \mathrm{TeV} \dv*{\sigma}{m}$ fit.

\subsection{Case III}

\begin{figure}[htbp]
	\centering
	\includegraphics[width=0.55\linewidth]{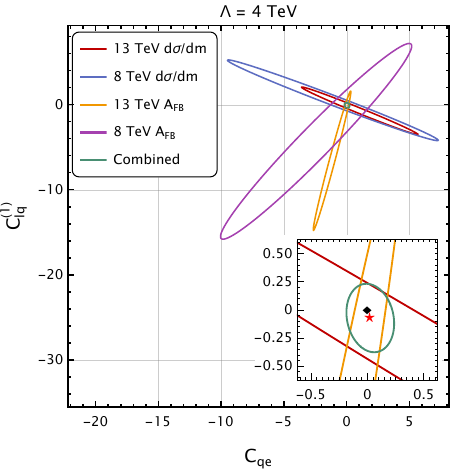}
	\caption{The 68\% C.L. bounds of the Wilson coefficients $C_{qe}$ and $C_{lq}^{(1)}$ from the $\chi^2$ fits (case III). The {blue, red, purple and orange} ellipses correspond to the fits with dataset {I, II, III} and IV (as labeled in Table~\ref{tab:datasets}), respectively. The green ellipse corresponds to the combined fit with all four datasets. The best-fit result is denoted with a red star, while the SM is denoted with a black diamond. }
	\label{fig:results:Cqe+Clq1}
\end{figure}

We now study two cases where flat directions are not present. We consider the case when $C_{qe}$ and $C_{lq}^{(1)}$ are nonzero. In the high invariant-mass limit, the flat directions exist separately in the up-quark and down-quark channels~\cite{Boughezal:2020uwq}. In the up-quark channel, the SMEFT contributions vanish under the following conditions after integrating over the angular variables:
\begin{align}
	C_{l q}^{(1)}=-C_{q e} \frac{Q_u e^2-g_Z^2 g_L^u g_R^e}{Q_u e^2-g_Z^2 g_L^e g_L^u} ,
\end{align}
while in the down-quark channel: 
\begin{align}
	C_{l q}^{(1)}=-C_{q e} \frac{Q_d e^2-g_Z^2 g_L^d g_R^e}{Q_u e^2-g_Z^2 g_L^e g_L^d}.
\end{align}
These two conditions cannot be simultaneously satisfied. Therefore, the Drell-Yan invariant-mass measurements are better suited to constraining $C_{qe}$ and $C_{lq}^{(1)}$ than in the previous cases. The 68\% C.L. ellipses of $C_{qe}$ and $C_{lq}^{(1)}$ are shown in Fig.~\ref{fig:results:Cqe+Clq1}. While the degeneracies between $C_{qe}$ and $C_{lq}^{(1)}$ are much weaker compared to the two cases above, the constraints still exhibit elongated shapes, although they do not extend as far as in the previous cases. However, the ellipses from $\dv*{\sigma}{m}$ and AFB measurements are almost orthogonal to each other. Consequently, the combined fit significantly improves the constraints.  

\begin{figure}[htbp]
	\centering
	\includegraphics[width=0.55\linewidth]{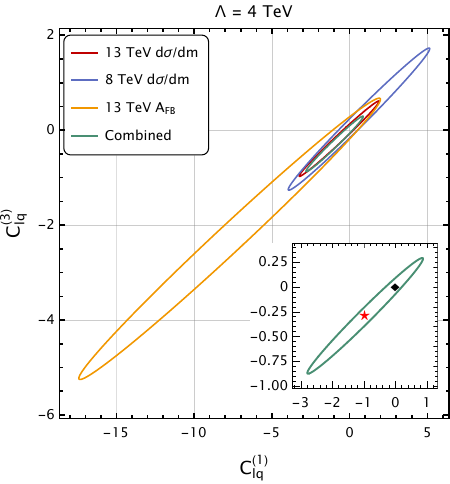}
	\caption{The 68\% C.L. bounds of the Wilson coefficients $C_{lq}^{(1)}$ and $C_{lq}^{(3)}$ from the $\chi^2$ fits (case IV). The {blue, red and orange} ellipses correspond to the fits with dataset I, II and IV (as labeled in Table~\ref{tab:datasets}), respectively. The green ellipse corresponds to the combined fit with all four datasets. {We note that the dataset {III} constraint is significantly weaker than the other three constraints, and is not shown on this plot. }The best-fit result is denoted with a red star, while the SM is denoted with a black diamond. }
	\label{fig:results:Clq1+Clq3}
\end{figure}

\subsection{Case IV}

Finally, we consider the case when $C_{lq}^{(1)}$ and $C_{lq}^{(3)}$ are nonzero. The Drell-Yan cross sections in the up-quark channel depend on the combination $C_{lq}^{(1)}-C_{lq}^{(3)}$, while in the down-quark channel they depend upon $C_{lq}^{(1)}+C_{lq}^{(3)}$. The 68\% C.L. ellipses of $C_{lq}^{(1)}$ and $C_{lq}^{(3)}$ are shown in Fig.~\ref{fig:results:Clq1+Clq3}. While there is no flat direction present, the ellipses are still slightly elongated. However, they do not extend nearly as far in the parameter space as the previous cases. Unlike case III, the ellipses from $\dv*{\sigma}{m}$ and AFB measurements exhibit similar correlations. As a result, the combined fit is not able to improve upon the constraints. 

{
\subsection{Marginalized constraints}
\begin{figure}[htbp]
	\centering
	\includegraphics[width=0.5\linewidth]{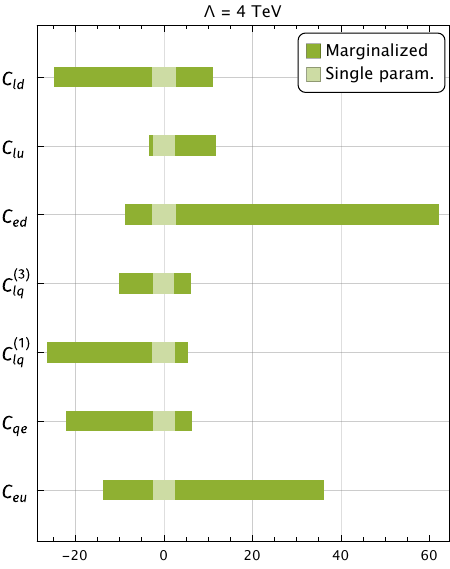}
	\caption{Marginalized 68\% C.L. bounds from all four datasets ($\Lambda=4\ \mathrm{TeV}$). Only one Wilson coefficient is enabled for the lighter green bars. For the darker ones, all seven Wilson coefficients are enabled and the other six are marginalized over. Only linear contributions of dimension-6 operators are included. \label{fig:MarginalizeDim6}}
\end{figure}

We next study the scenario where all Wilson coefficients are turned on. In Fig.~\ref{fig:MarginalizeDim6}, we show the 68\% C.L. bounds for each dimension-6 four-fermion Wilson coefficient while marginalizing over all six other operators. While the degeneracies are particularly severe for the 13 TeV AFB dataset, the combined fit is able to significantly improve the constraints for most of the operators. Nevertheless, the combined fits become worse after marginalizing over the other operators. As observed in the previous cases, the AFB datasets can only reduce degeneracies in some cases. Thus, after marginalizing over Wilson coefficients, the combined fits are not as strong as the single-parameter fits. 
}

\section{Impact of quadratic dimension-6 terms\label{sec:quadratic}}

In this section, we study the impact of the quadratic corrections in the dimension-6 SMEFT Wilson coefficients on the fits. As shown in Eq.~\eqref{eq:xsection}, the quadratic corrections $b_{ij}$ contribute at $\mathcal{O}(1/\Lambda^4)$, while the linear corrections $a_i$ contribute at $\mathcal{O}(1/\Lambda^2)$. Although the quadratic corrections are suppressed by $\Lambda^2$, they are not negligible. In the case of strong degeneracies, the linear corrections cancel in certain directions, making the quadratic corrections important. 

\begin{figure}[htbp]
	\centering
	\includegraphics[width=0.55\linewidth]{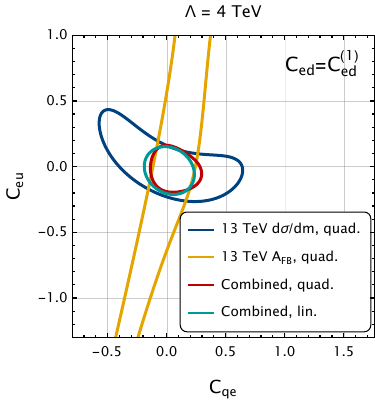}
	\caption{The 68\% C.L. bounds of the Wilson coefficients $C_{e u}$ and $C_{qe}$ from the $\chi^2$ fits with the quadratic terms included. The blue and orange ellipses correspond to the fits with dataset II and IV (as labeled in Table~\ref{tab:datasets}), respectively. The red (green) ellipse corresponds to the combined fit with all four datasets and with (without) quadratic terms. }
	\label{fig:results:Cqe+Ceu_quad}
\end{figure}

\begin{figure}[htbp]
	\centering
	\includegraphics[width=0.55\linewidth]{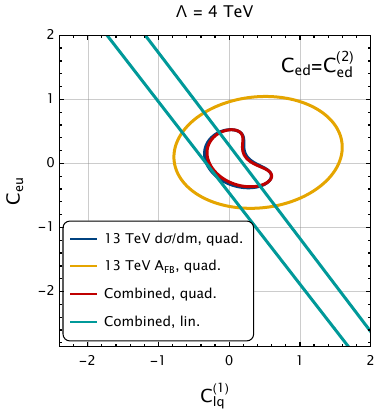}
	\caption{The 68\% C.L. bounds of the Wilson coefficients $C_{e u}$ and $C_{lq}^{(1)}$ from the $\chi^2$ fits with the quadratic terms included. The blue and orange ellipses correspond to the fits with dataset II and IV (as labeled in Table~\ref{tab:datasets}), respectively. The red (green) ellipse corresponds to the combined fit with all four datasets and with (without) quadratic terms. }
	\label{fig:results:Clq1+Ceu_quad}
\end{figure}

We take case I as a first example. In Fig.~\ref{fig:results:Cqe+Ceu_quad}, we show the 68\% C.L. bounds on $C_{e u}$ and $C_{qe}$ from the $\chi^2$ fits with the quadratic terms included. The individual ellipses with $\dv*{\sigma}{m}$ and AFB measurements still exhibit slightly elongated shapes, but the significant degeneracies found in Fig.~\ref{fig:results:Cqe+Ceu} are no longer present. The combined fit produces a much more circular ellipse, and the constraints are improved in the individual fits with quadratic terms. As discussed earlier it is not generally possible to simultaneously set the $A$ and $B$ terms to zero in Eq.~(\ref{eq:DYcr}) with non-trivial values of the Wilson coefficients. The quadratic terms therefore break the degeneracies found in the separate invariant-mass and AFB linear fits. However, we observe that the combined fit does not show much difference after including quadratic terms. Because of the degeneracies already being removed by including both $\dv*{\sigma}{m}$ and AFB data, the combined linear fit is nearly identical to the combined quadratic fit.

With case II, we observe that the inclusion of quadratic terms has much more impact on the fits. In Fig.~\ref{fig:results:Clq1+Ceu_quad}, we show the 68\% C.L. bounds of $C_{e u}$ and $C_{lq}^{(1)}$ from the $\chi^2$ fits with the quadratic terms included. As discussed earlier, the invariant-mass and AFB datasets probe the same linear combination of Wilson coefficients for case II when the SMEFT expansion is truncated at $1/\Lambda^2$, leading to the observed degeneracy in the combined linear fit. This is broken by the inclusion of terms quadratic in the Wilson coefficients.
 
\begin{figure}[htbp]
	\centering
	\includegraphics[width=0.55\linewidth]{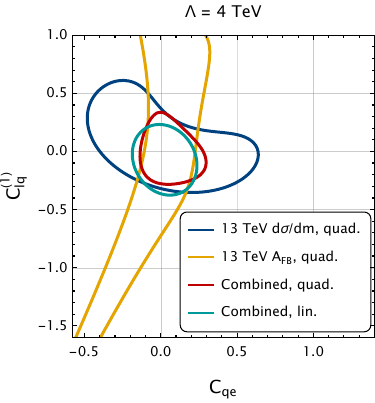}
	\caption{The 68\% C.L. bounds of the Wilson coefficients $C_{qe}$ and $C^{(1)}_{lq}$ from the $\chi^2$ fits with the quadratic terms included. The blue and orange ellipses correspond to the fits with dataset II and IV (as labeled in Table~\ref{tab:datasets}), respectively. The red (green) ellipse corresponds to the combined fit with all four datasets and with (without) quadratic terms. }
	\label{fig:results:Clq1+Ceq_quad}
\end{figure}

\begin{figure}[htbp]
	\centering
	\includegraphics[width=0.55\linewidth]{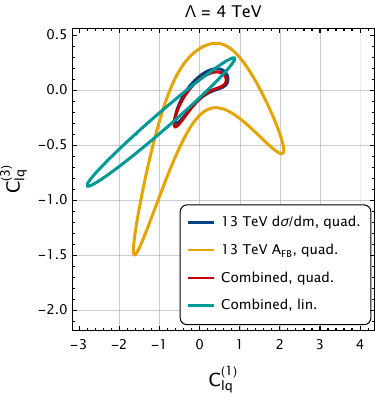}
	\caption{The 68\% C.L. bounds of the Wilson coefficients $C^{(1)}_{lq}$ and $C^{(3)}_{lq}$ from the $\chi^2$ fits with the quadratic terms included. The blue and orange ellipses correspond to the fits with dataset II and IV (as labeled in Table~\ref{tab:datasets}), respectively. The red (green) ellipse corresponds to the combined fit with all four datasets and with (without) quadratic terms. }
	\label{fig:results:Clq1+Clq3_quad}
\end{figure}

Finally, we present the comparison between the linear and quadratic fits for cases III and IV in Figs.~\ref{fig:results:Clq1+Ceq_quad} and~\ref{fig:results:Clq1+Clq3_quad}, respectively. We recall that in both of these cases we cannot solve analytically for a flat direction in the high-energy limit because of the structure of the SMEFT corrections to the Drell-Yan cross section. However, in case IV we observed that AFB and the invariant-mass distribution exhibited similar correlations between the Wilson coefficients, and not much was gained by combining the two measurements. This was not the situation for case III, where the measurements exhibited distinct correlations. The impact of this is seen in the quadratic fits. There is good agreement between the linear and quadratic fits for case III, since including the AFB measurement in the fit has already cut off the slight elongation that occurs with the invariant-mass data alone. For case IV there is a difference, since in this case AFB does not add to the invariant-mass measurements.

{
	When considering the $\mathcal{O}(1/\Lambda^4)$ term in the SMEFT expansion, the effect of the dimension-8 operators should be considered. This was done for the Drell-Yan invariant-mass distribution in~\cite{Boughezal2021}. We extend that analysis here by studying the effect of more experimental datasets and the impact of AFB on the fits. We choose the pairs ($C_{eu}$, $C^{(2)}_{e^2u^2D^2}$) and ($C_{lu}$, $C^{(2)}_{l^2u^2D^2}$) as representative examples. The dimension-8 operators corresponding to the Wilson coefficients $C^{(2)}_{e^2u^2D^2}$ and $C^{(2)}_{l^2u^2D^2}$ were shown to lead to a non-trivial angular dependence in the Drell-Yan process~\cite{Alioli:2020kez} that can be probed by $A_{FB}$ measurements. We note that the helicity structure of the fermions is the same for both the dimension-6 and dimension-8 coefficients in both pairs. This is what would be obtained in a UV model where the effective operators are obtained by integrating out a heavy resonance that only couples to a specific helicity combination. The 68\% C.L. bounds for these two Wilson coefficient pairs are shown in Fig. 15. These two pairs illustrate the possibilities that occur when studying the interplay between dimension-6 and dimension-8 coefficients. Due to the strong correlation between dimension-6 and dimension-8 effects for LHC energies, the dimension-8 operators have significant impact on the constraints of dimension-6 Wilson coefficients when the invariant-mass or AFB distributions are considered separately. This can be seen from comparing the ellipses for the individual datasets, which denote the marginalized constraints, with the colored bands, which denote the constraints when only a single Wilson coefficient is activated. However, in the  ($C_{eu}$, $C^{(2)}_{e^2u^2D^2}$)  the combination of AFB and invariant-mass measurements breaks this degeneracy and the marginalized constraints approach the individual Wilson coefficient constraints. For the second case, however, both $\dv*{\sigma}{m}$ and AFB measurements exhibit similar correlations, and the inclusion of AFB measurements does not improve the constraints.
}

\begin{figure}[htbp]
	\centering 
	\includegraphics[width=0.49\linewidth]{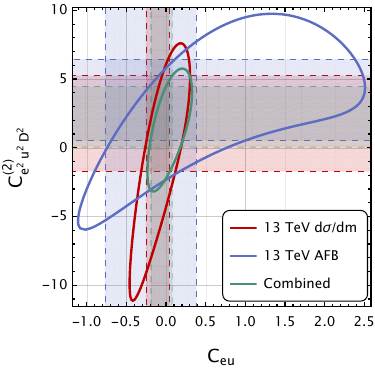}
	\includegraphics[width=0.49\linewidth]{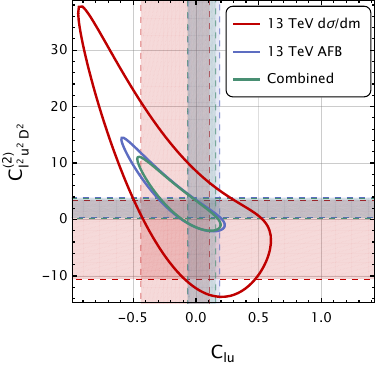}
	\caption{The 68\% C.L. bounds of one dimension-6 and one dimension-8 Wilson coefficients. The left panel shows the ellipses of $C_{eu}$ and $C_{e^2u^2D^2}^{(2)}$, while the right panel shows the ellipses of $C_{lu}$ and $C_{l^2u^2D^2}^{(2)}$. The red and blue lines correspond to the fits with dataset II and IV (as labeled in Table~\ref{tab:datasets}), respectively. The green line corresponds to the combined fit with all four datasets. The shaded areas enclosed by dashed lines represent 1D fit with only one Wilson coefficient enabled, while the ellipses represent 2D fits with both Wilson coefficients enabled. \label{fig:results:dim8}}
\end{figure}

\begin{figure}[htbp]
	\centering
	\includegraphics[width=0.7\linewidth]{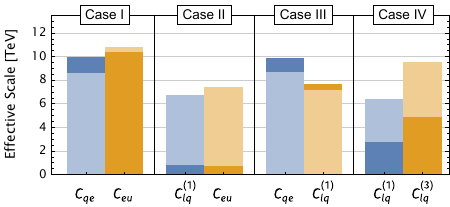}
	\caption{Effective scales for the Wilson coefficients obtained from the 68\% C.L. bounds. The bars with lighter colors correspond to the fits with quadratic terms included, while the bars with darker colors correspond to the fits without quadratic terms. }
	\label{fig:results:effscale}
\end{figure}

	In Fig.~\ref{fig:results:effscale}, we show the effective scales for the Wilson coefficients obtained from our $\chi^2$ fits. We recall that the effective scale $M$ connects parameters in the ultraviolet completion with the SMEFT parameters:
	\begin{align}
		\frac{C}{\Lambda^2} &\sim \frac{g^2}{M^2} \label{eq:effscale} . 
	\end{align}
	We assume the coupling strength $g$ in UV models to be unity, and $\Lambda=4\ \mathrm{TeV}$. Therefore, the effective scale $M$ becomes
	\begin{align}
		M &= \frac{\Lambda}{\sqrt{\abs{C}}} \label{eq:effscale2} .
	\end{align}

	We observe that due to severe degeneracies between Wilson coefficients without the quadratic terms, the effective scales probed are significantly lower in cases II and IV. In the extreme scenario of case II, where strong degeneracies are present even after combining invariant-mass and AFB data, the effective scales are lower than $1\ \mathrm{TeV}$. These degeneracies are removed by the inclusion of the quadratic terms, and we observe a visible increase of effective scales for cases II and IV after including the quadratic terms. Since the degeneracies are already removed by combining the $\dv*{\sigma}{m}$ and the AFB measurements in cases I and III, the inclusion of quadratic terms has little impact on the effective scales for these cases. {We note that a detailed study of constraints on the coefficients $C_{qe}$, $C_{eu}$, $C_{lq}^{(1)}$ and $C_{lq}^{(3)}$ in the MFV scenario from flavor-violating measurements was performed in~\cite{Aoude2020}. The flavor constraints were found there to significantly improve upon bounds from non-flavor violating measurements such as $W$-pair production and precision $Z$-pole observables from LEP only for the coefficient $C_{lq}^{(3)}$.}

\section{Conclusions\label{sec:conclusion}}

In this work, we have investigated how AFB measurements at the LHC contribute to fits of the semileptonic four-fermion sector of the SMEFT. One issue identified in previous work is that invariant-mass measurements alone exhibit significant degeneracies in the Wilson coefficient parameter space. A main goal of this work was to learn in what instances AFB measurements can resolve these blind spots. We have studied constraints from joint fits to high-energy, high-luminosity datasets, and have provided a detailed description of our treatment of the experimental data. In most cases the combined fits impose stringent multi-TeV bounds on the parameter space since, and the inclusion of AFB data drastically improves the fit compared to using invariant-mass data alone. In some cases, however, we found that AFB measurements do not break the degeneracies, and do not improve the fits. This highlights the need for future datasets such as those from the EIC in order to fully probe this sector of the SMEFT. We have studied the impact of quadratic dimension-6 terms on the fits. In cases where AFB resolves the blind spots in the parameter space, the linear and quadratic fits are generally in good agreement. In other cases where the AFB and invariant-mass measurements exhibit similar correlations there are significant differences between the linear and quadratic fits, indicating the need to pay close attention to the convergence of the SMEFT expansion.

\begin{acknowledgments}
	We thank O.~Amram for helpful communication regarding the CMS result in~\cite{CMS:2022uul}.
		R.~B. is supported by the DOE contract DE-AC02-06CH11357.  Y.~H and F.~P. are supported by the DOE grants DE-FG02-91ER40684 and DE-AC02-06CH11357.
\end{acknowledgments}

\appendix 
\section{Details of experimental binning}\label{app}

The binning adopted for each dataset is listed in Table~\ref{tab:binning}. Note that all datasets have a dilepton rapidity cut $\abs{y_{ll}}\leq2.4$ except for dataset II. dataset III additionally bins in $\abs{y_{ll}}$. dataset I has an explicit description of the transverse momentum $p_T$ and pseudorapidity $\eta$ cuts on the leptons. The cuts are $p_T^{\ell_1}>40$ GeV, $p_T^{\ell_2}>30$ GeV, $\abs{\eta^{\ell_1}}<2.5$, $\abs{\eta^{\ell_2}}<2.5$.

{\setlength{\tabcolsep}{6pt}  
\begin{table}[htbp]
  \centering
  \begin{tabular}{c|p{10cm}cc}
    \hline
    \hline
    dataset           & $m_{ll}$ edges $[\mathrm{GeV}]$ & $\abs{y_{ll}}$ edges\\
    \hline
    I             & [116, 130, 150, 175, 200, 230, 260, 300, 380, 500, 700, 1000, 1500]  &[0.,2.4]      \\
    II ($ee$)     & [200, 220, 240, 260, 280, 300, 320, 340, 360, 380, 400, 
											420, 440, 460, 480, 500, 520,540, 560, 580, 600, 630, 660, 690, 720, 750, 780, 810, 840, 870, 900, 950, 1000, 1050, 1100, 1150, 1200, 1250, 1310, 1370, 1430, 1490, 1550, 1680, 1820, 1970, 2210, 6070]    &[0,$\infty$]    \\
    II ($\mu\mu$) & [209.63, 227.02, 245.86, 266.25, 288.34, 312.26, 338.16, 366.21, 396.59, 429.49, 465.12, 503.71, 545.49, 590.74, 639.75, 692.82, 750.29, 812.54, 879.94, 952.94, 1032., 1117.6, 1210.3, 1310.7, 1419.4, 1537.2, 1664.7, 1802.8, 1952.4, 2289.7, 7000.]    &[0,$\infty$]    \\
    III           & [120, 133, 150, 171, 200, 320, 500, 2000]    &[0., 1., 1.25, 1.5, 2.4]    \\
    IV            & [170, 200, 250, 320, 510, 700, 1000, 13000]    &[0.,2.4]    \\
    \hline
    \hline
  \end{tabular}
	\caption{The $m_{ll}$ and $y_{ll}$ edges for each dataset. The first column is the indexed number of the datasets, as shown in Tab~\ref{tab:datasets}. The binning in the electron and muon channels of dataset II are taken differently. The second column is the $m_{ll}$ edges, and the third column is the $y_{ll}$ edges.}
	\label{tab:binning}
\end{table}}

\bibliography{./SMEFT.bib}
\end{document}